\documentclass[12pt]{iopart}
\usepackage{epsf,amssymb}

\renewcommand{\thefootnote}{\arabic{footnote}}
\def\etal{{\em et.al.}}

\def\figFourFlGr
{
\begin{figure}[htb]
 \epsfxsize=0.8\textwidth
 \centerline{\epsfbox{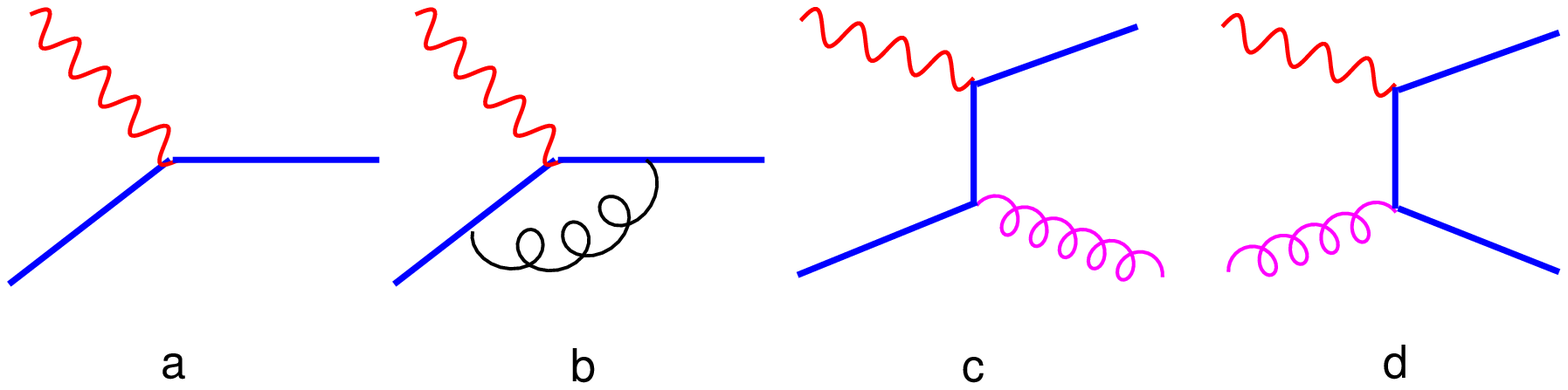}}
 \caption{Partonic processes for charm production to NLO in the 4-flavor scheme.}
 \label{fig:FourFlGr}
\end{figure}
}
\def\figCartoon
{
\begin{figure}[htb]
 \centerline{
 (a)
  \epsfxsize=0.4\textwidth
  \epsfbox{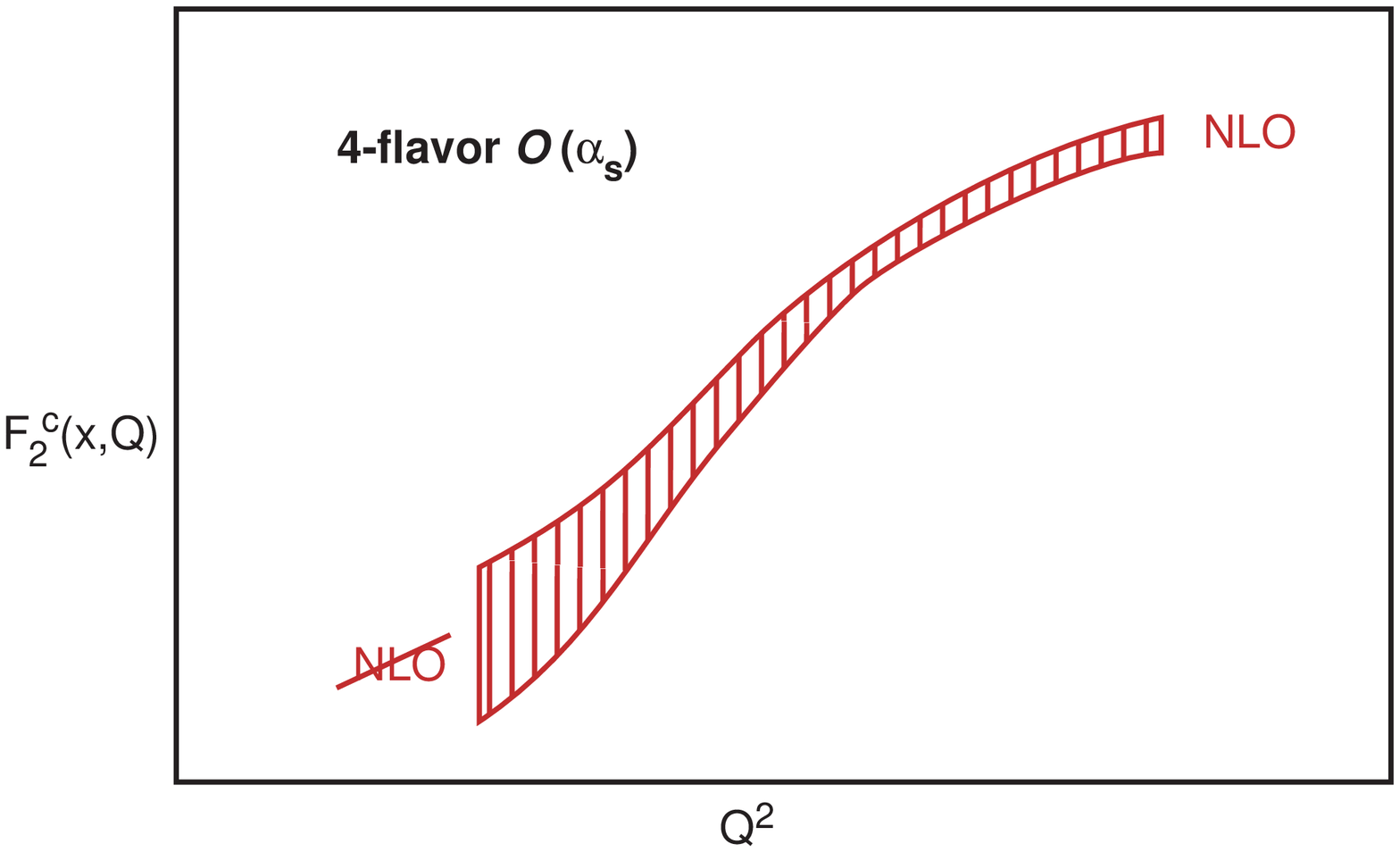}
 (b)
  \epsfxsize=0.4\textwidth
  \epsfbox{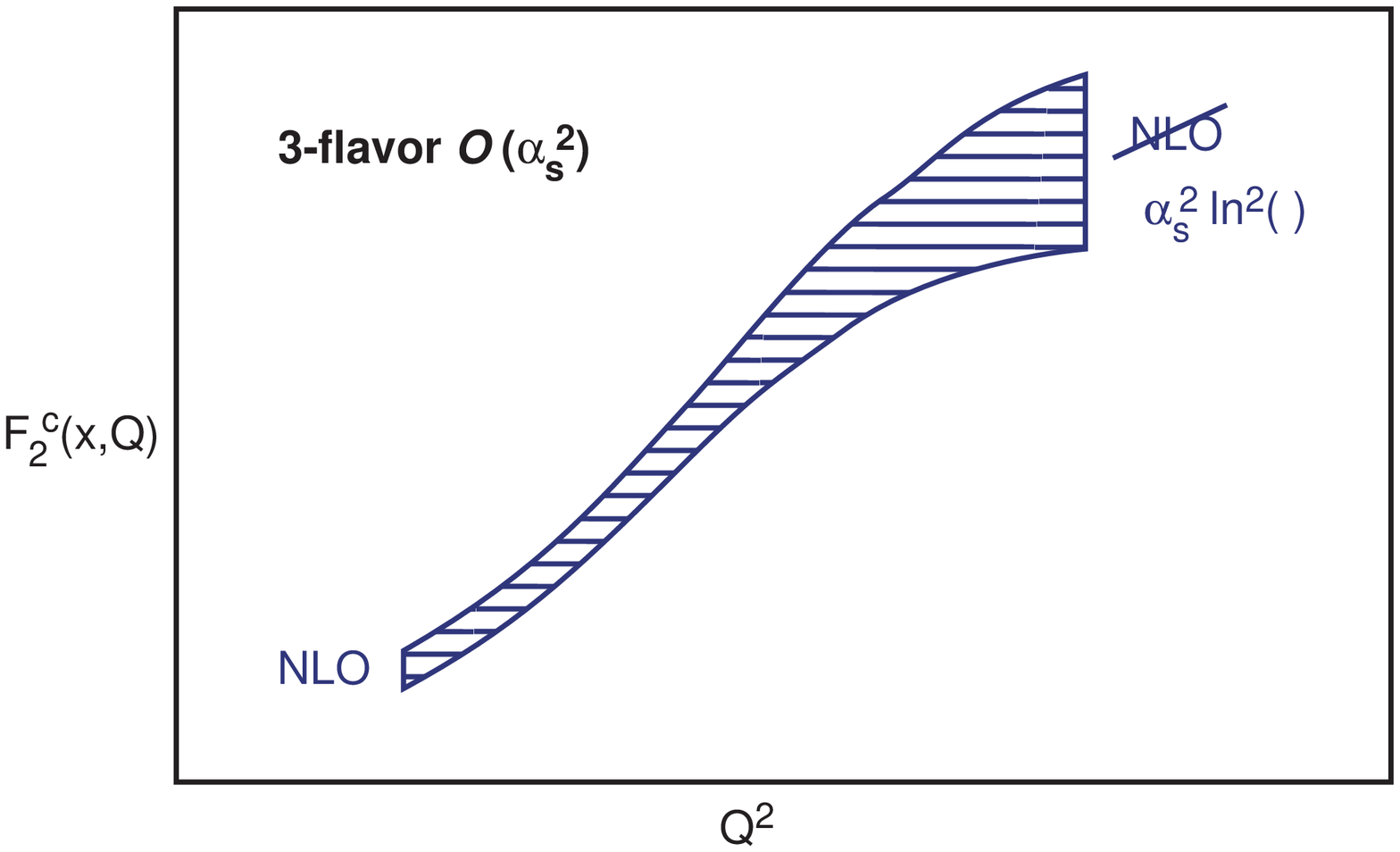}
 }
 \caption{Expected regions of applicability and uncertainty of the 4-flavor (a)
 and 3-flavor (b) schemes. Note: (i) the power of $\alpha_s$ for
 ``NLO'' is different in the two schemes due to the resummation of perturbation series;
 and (ii) the reliability of the calculation in each scheme depends on the scale $Q$.}
 \label{fig:Cartoon}
\end{figure}
}
\def\figThreeFlGr
{
\begin{figure}[htb]
 \epsfxsize=0.8\textwidth
 \centerline{\epsfbox{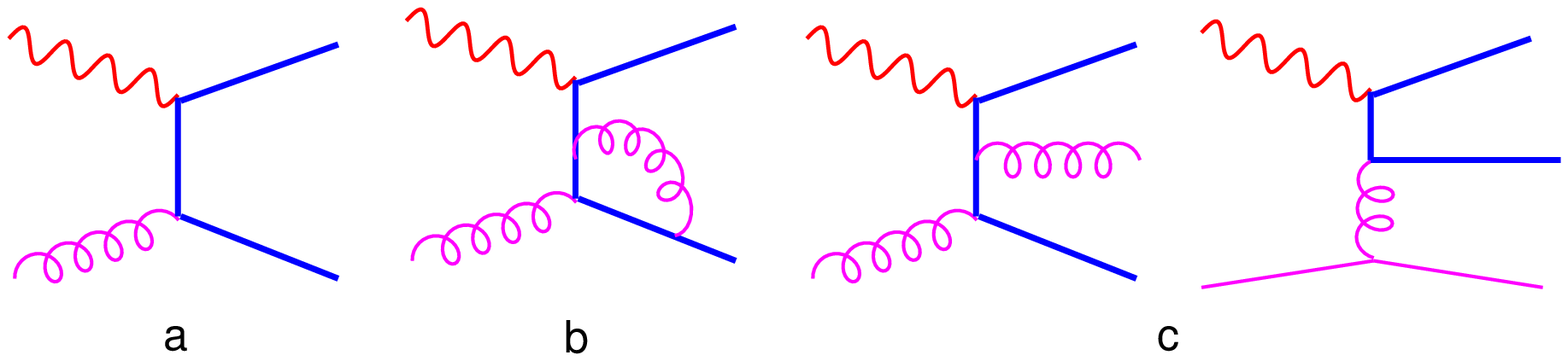}}
 \caption{Partonic processes for charm production to NLO in the 3-flavor scheme.}
 \label{fig:ThreeFlGr}
\end{figure}
}
\def\figFive
{
\begin{figure}[htb]
 \centerline{(a)
 \epsfxsize=0.27\textwidth \epsfbox{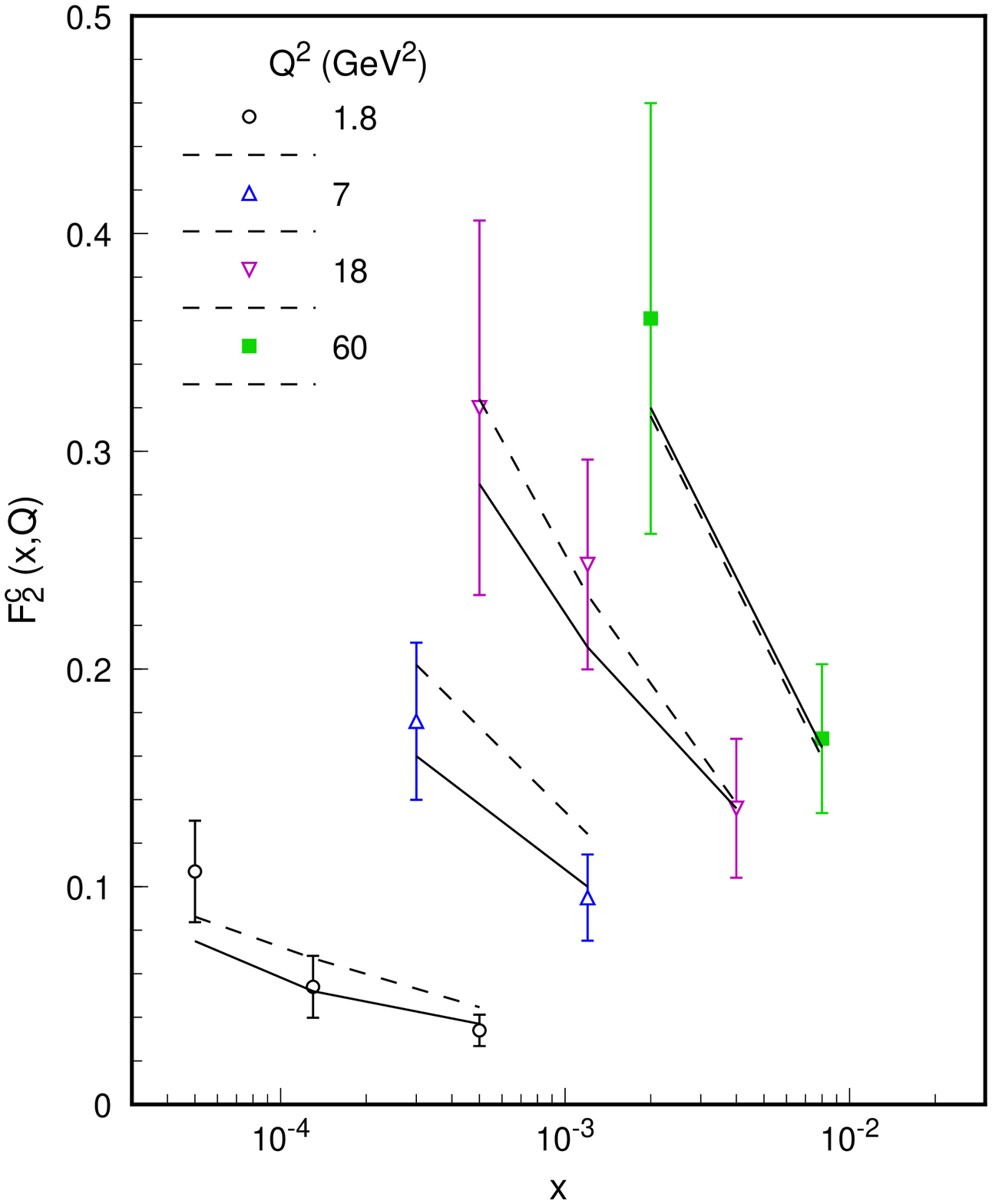}
 \epsfxsize=0.27\textwidth \epsfbox{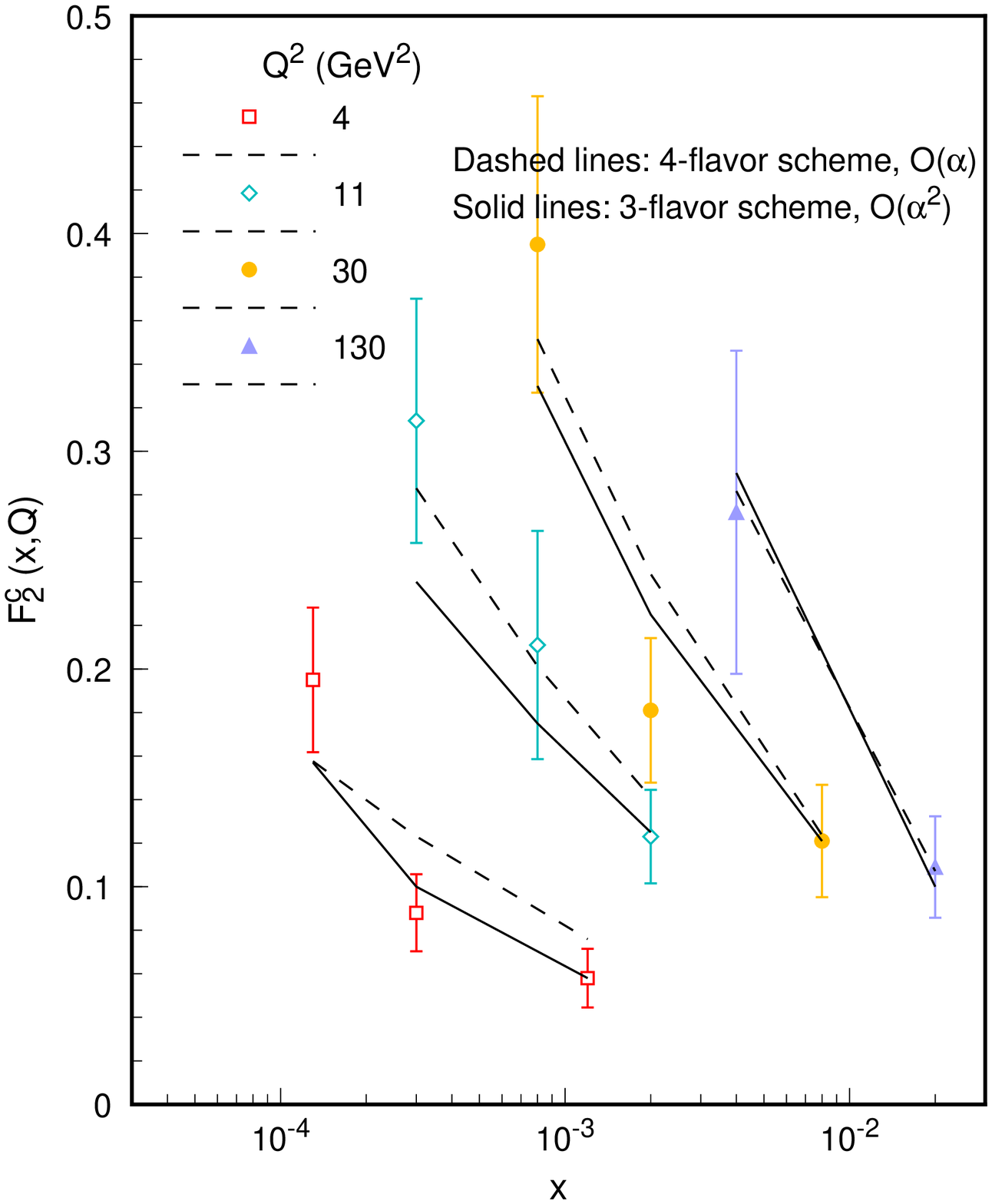}
 (b)}
 \caption{
Comparison of the inclusive charm production data of Zeus \cite{ZeusF2c}
with: (i) order $\alpha_s^2$ 3-flavor (NLO) calculation (solid lines); and
order $\alpha_s$ 4-flavor (also NLO) calculation ($m_c \neq 0$) in the
general formalism (dashed lines). The various $Q$ bins are alternately put
into two separate plots to avoid overlapping points and curves. Both schemes
appear to be robust within the experimental kinematic range.}
\label{fig:Five} \end{figure} } %

\def\figResult
{ \begin{figure}[htb]
 \centerline{(a)
 \epsfxsize=0.4\textwidth  \epsfbox{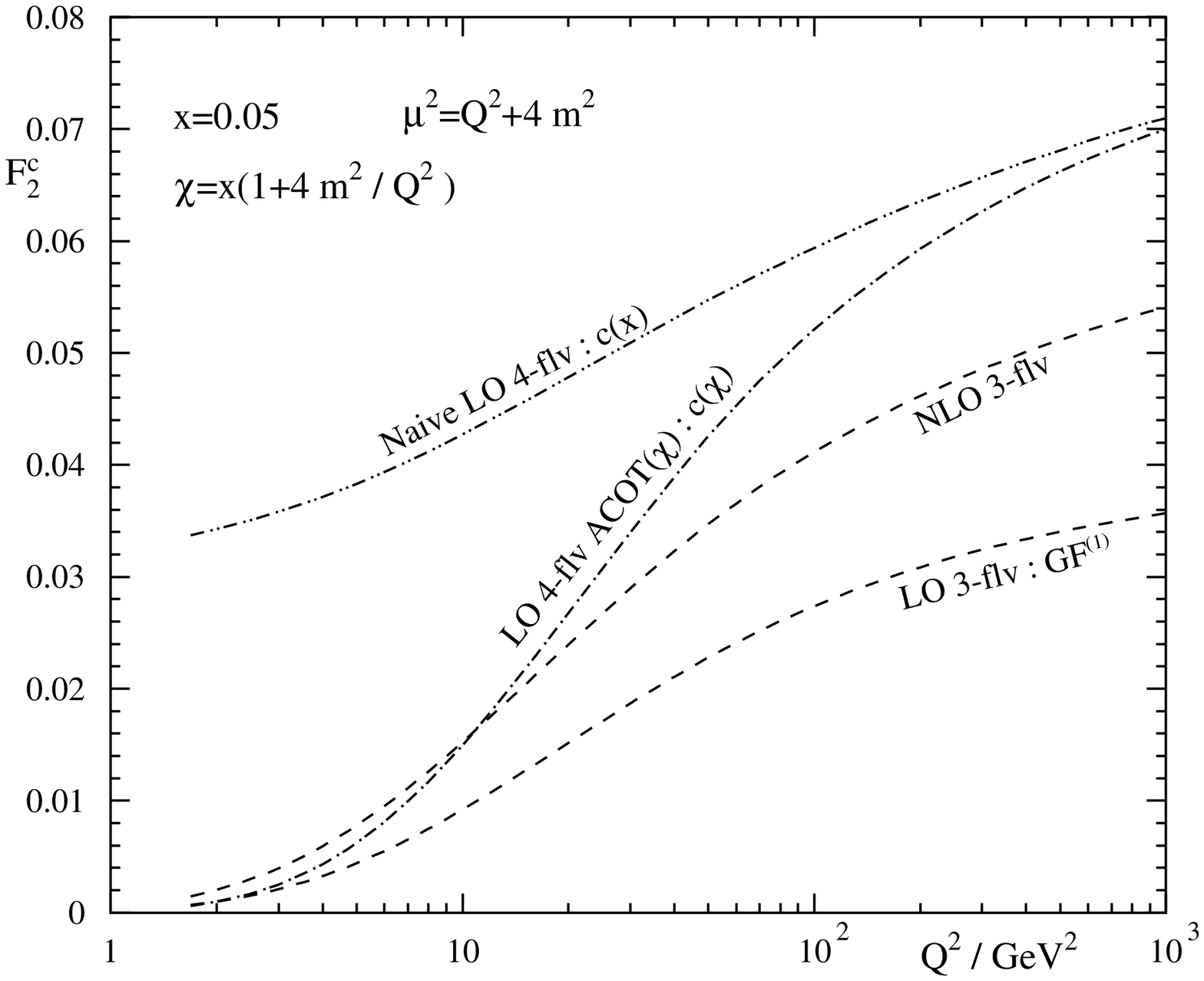}
 \hfill (b)
 \epsfxsize=0.4\textwidth  \epsfbox{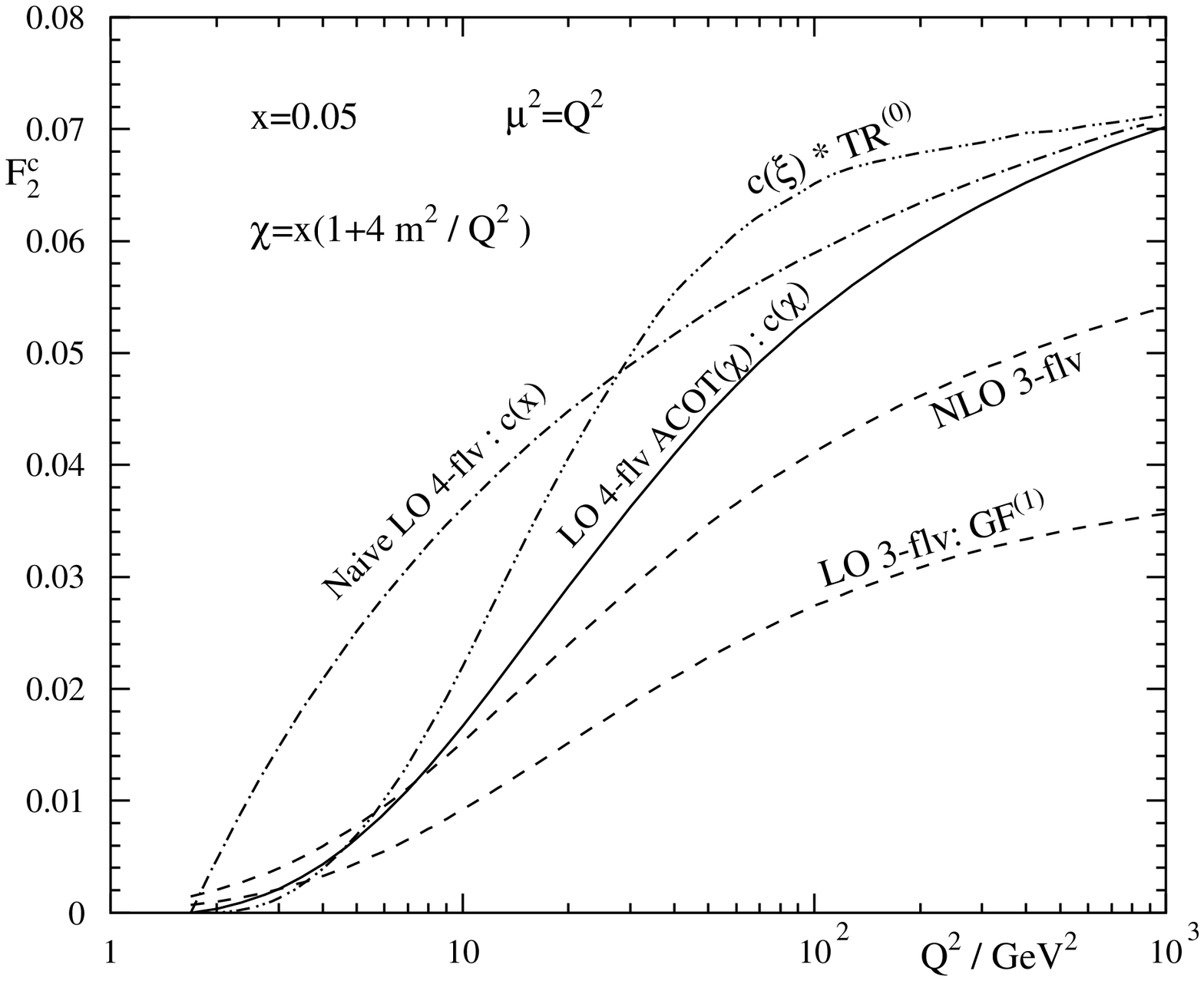}
 }
 \caption{Comparison of $F_2^c$ calculations vs.\ $Q^2$ with two different choices of
 $\mu$. The curve labelled ``$c(\xi))$ TR'' in (b), is the LO result using the
 prescription of Thorne-Roberts \cite{ThorneRob}.
 }
 \label{fig:Result}
\end{figure}
}
\def\figResX
{
\begin{figure}[htb]
 \epsfxsize=\textwidth  \epsfbox{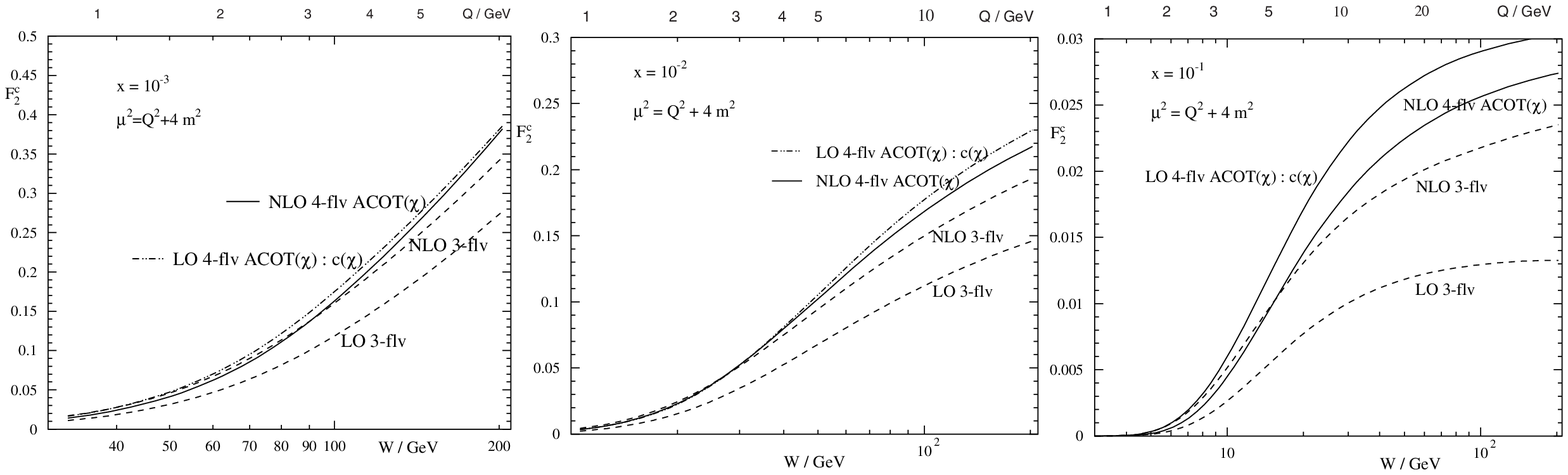}
 \caption{$F_2^c$ vs.\ $W$ ($Q$) for three $x$ values, to give an overall view
 of the behavior of the LO and NLO ACOT($\chi$) calculations, and their comparison
 to the LO and NLO results of the 3-flavor scheme.  See discusions in the text.}
 \label{fig:ResX}
\end{figure}
}

\begin{document}

%\Input{cover.tex}
\title{Open Heavy Flavor Production in QCD -- Conceptual Framework and
Implementation Issues}

\author{Wu-Ki Tung, Stefan Kretzer, Carl Schmidt
\footnote[3]{Presented by Wu-Ki Tung at Ringberg Workshop: \emph{New Trends
in HERA Physics 2001}, Munich, Germany. E-mail: Tung@pa.msu.edu.} }
\address{Department of Physics/Astronomy, Michigan State University}

\begin{abstract}
Heavy flavor production is an important QCD process both in its own right and
as a key component of precision global QCD analysis.  Apparent disagreements
between fixed-flavor scheme calculations of b-production rate with
experimental measurements in hadro-, lepto-, and photo-production provide new
impetus to a thorough examination of the theory and phenomenology of this
process. We review existing methods of calculation, and place them in the
context of the general PQCD framework of Collins. A distinction is drawn
between scheme dependence and implementation issues related to quark mass
effects near threshold. We point out a so far overlooked kinematic constraint
on the threshold behavior, which greatly simplifies the variable flavor
number scheme. It obviates the need for the elaborate existing prescriptions,
and leads to robust predictions. It can facilitate the study of current
issues on heavy flavor production as well as precision global QCD analysis.

\end{abstract}



%\maketitle

\renewcommand{\thefootnote}{\arabic{footnote}}\setcounter{footnote}{0}
%\Input{intro.tex}
%TCIDATA{LaTeXparent=0,0,RingbergProc.tex}

%TCIDATA{ChildDefaults=chapter:1,page:1}

\section{Introduction}
\label{sec:Intro}

Conventional perturbative Quantum Chromodynamics (PQCD) theory is formulated
most simply in terms of zero-mass partons. For processes depending on only
one hard scale $Q,$ the well-known factorization theorem provides a
straightforward procedure for order-by-order perturbative calculations, as
well as an associated intuitive parton picture interpretation of the
perturbation series. Heavy quark production presents a challenge in PQCD
because the heavy quark mass, $m_{H}$ $(H=c,b,t),$ provides an additional
hard scale which complicates the situation.  The perturbative series must be
organized in different ways depending on the relative magnitudes of $m_{H}$
and $Q$.

A reliable formulation of heavy quark production is important for high energy
physics because of its intrinsic value as a fundamental process, as well as
because of its significant contribution to total inclusive cross-sections at
high energies. Recent heightened interest has been directed to this process
on both of these accounts. First, evidence is accumulating from several
processes in hadro- \cite{CDFb,D0b}, photo- \cite{ZeusPhB}, lepto-, and
$\gamma\gamma$-production of bottom \cite{Sefkow:2001bs}, that experimental
cross-sections are uniformly larger than existing calculations, by roughly a
factor of 2. This is in contrast to charm production where no such gross
discrepancy is seen. Secondly, the global QCD analysis of the very accurately
measured deep inelastic scattering structure functions now definitely demands
a quantitative treatment of charm production which incorporates heavy quark
mass effects in a reliable way.

We first briefly review the various approaches to heavy quark production
\cite{pAcot,ACOT,Kniehl:1995em,BuzaEtal,ThorneRob,CaccGreNas}, and describe
their synthesis in the general framework of Collins \cite{Collins97}. This
provides a context to distinguish between issues relating to the choice of
factorization scheme and those relating to implementation prescriptions
allowed within a given scheme. The choice of prescriptions, such as that
associated with mass threshold behavior, can matter in physical applications.
In principle, the possible choices may be equivalent within the PQCD
formalism; but in practice, some are natural and robust, while others appear
to be more ad hoc and volatile. We examine this practical problem, and point
out a so far overlooked kinematic constraint which greatly simplifies the
calculation of subprocesses with heavy quark initial states. This leads to a
efficient formalism which also yields very robust predictions.  It can be
used to address the current challenges described in the previous paragraph.

%\Input{background.tex}
%TCIDATA{LaTeXparent=0,0,RingbergProc.tex}

%TCIDATA{ChildDefaults=chapter:1,page:1}

\section{Conventional Approaches}
\label{sec:ConvAppr}

To see the basic physics ideas, let us focus explicitly on the production of
charm ($H=c$) in deep inelastic scattering (DIS). All considerations apply to
a generic heavy quark, and to hadro-production processes.  The two standard
methods for PQCD calculation of heavy quark processes represent two
diametrically opposite ways of reducing the two-scale problem to an effective
(hence approximate) one-scale problem.

\subsection{4-flavor Zero-mass Scheme}

In the conventional \underline{parton-model approach}, used in many global
QCD analyses of parton distributions (e.g.\ {\small EHLQ, MRS, CTEQ}) and
Monte Carlo programs (e.g.\ {\small ISAJET, PYTHIA, HERWIG}), the textbook
zero-mass parton approximation is applied to a heavy
quark calculation as soon as the typical energy scale of the physical process $Q$%
\footnote{%
We use $Q$ as the generic name for a typical kinematic physical scale. It
could
be $Q$, $W$, or $p_{T}$, depending on the process.} %
is above the mass scale $m_{c}$. This leaves $Q$ as the only apparent hard
scale in the problem. The LO and NLO production
mechanisms for charm are given by Fig.~\ref{fig:FourFlGr},%
\figFourFlGr%
where the solid lines represent the charm quark. Note that the NLO diagrams
are of order $\alpha_s$ , just as for the familiar case of total inclusive
DIS structure functions.  This is the most natural calculational scheme to
adopt at high energies when $Q\gg m_{c}$. However, as we go down the energy
scale toward the charm production threshold ($W=2m_c$) region, it becomes
unreliable because the approximation $m_{c}=0$ deteriorates as $Q$ becomes of
the same order of magnitude as $m_{c}$. In this limit, all terms in
Fig.~\ref{fig:FourFlGr} become effectively of order $\alpha_s$ (assuming no
non-perturbative charm), the formal ``NLO'' designation losses meaning. This
point is illustrated qualitatively in Fig.~\ref{fig:Cartoon}a as an
uncertainty band marked by vertical hashes which is narrow at large $Q$ but
is expected to widen as $Q$ decreases.
%In addition, the approximate treatment of
%kinematics breaks down near threshold, as will be discussed in detail later.%
\figCartoon

\subsection{3 (Fixed)-Flavor Scheme}

In the \underline{heavy quark approach} which played a dominant role in ``NLO
calculations'' of the production of heavy quarks \cite{FFN}, the quark is
always treated as a ``heavy'' particle and never as a parton. The mass
parameter $m_{c}$ is explicitly kept along with $Q$ as if they are of the
same order, \emph{irrespective of their real relative magnitudes}. This is
usually referred to as the fixed flavor-number (FFN) scheme. The LO and NLO
partonic processes in this scheme are exemplified by the type of diagrams
shown in Fig.~\ref{fig:ThreeFlGr}. In this case, the NLO diagrams are of
order $\alpha_s^2$, which are much more complicated to calculate. Near
threshold $W=2m_c$, it is natural to consider the charm quark as a heavy
particle, hence the NLO calculation in this scheme is reliable (unless there
is non-perturbative charm). However, as $Q$ becomes large compared to
$m_{c}$, the FFN approach becomes unreliable since the perturbative expansion
contains terms of the form $\alpha_{s}^{n}\log ^{n}\left(
m_{c}^{2}/Q^{2}\right) $ at any order $n$, which ruin the convergence of the
series. These terms are \emph{not infra-red safe} as $m_{c}$ $\rightarrow 0$
or $Q\rightarrow \infty $. Furthermore, the calculation is no longer NLO in
accuracy, in spite of the hard $\mathcal{O}(\alpha_s^2)$ calculation when
$\alpha_{s}\log(m_{c}^{2}/Q^{2})\sim 1 $. This is illustrated in
Fig.~\ref{fig:Cartoon}b as an uncertainty band marked by horizontal hashes
which is narrow near threshold but is expected to widen as $Q/m_{c}$ increases.%
\figThreeFlGr %

PQCD does not predict at what energy scale $Q$ do the large logarithm terms
actually become a problem \cite{BuzaEtal,GRS}. In practice, the 3-flavor
scheme has worked well in charm production phenomenology so far
\cite{Redondo:2001iy}. However, for reasons touched upon in the introduction
(and to be discussed in detail in subsequent sections), it is important to
examine the general picture.

%\Input{genframe.tex}
%TCIDATA{LaTeXparent=0,0,RingbergProc.tex}

%TCIDATA{ChildDefaults=chapter:2,page:1}

\section{The Unified Framework of Collins -- Factorization with massive partons}
\label{sec:GenFrame}

It is obvious from Figs.~\ref{fig:Cartoon}a,b that the two conventional
approaches are individually unsatisfactory over the full energy range, but
are mutually complementary. Therefore, the most reliable PQCD prediction for
the physical $F_{2}(x,Q)$ overall, can be obtained by combining the two,
utilizing the most appropriate scheme at a given energy scale $Q$. This
results in a composite scheme, as represented by the cross-hashed region in
Fig.\ 4, which is simply a composite of the two figures of
Fig.~\ref{fig:Cartoon}. The use of a composite scheme consisting of different
numbers of flavors in different energy ranges, rather than a fixed number of
flavors, is familiar in the conventional zero-mass parton picture. The new
formalism espoused in Refs.\ \cite{pAcot,ACOT} provides a quantum field
theoretical basis \cite{Collins97} for this\\
\parbox{0.6\textwidth}{
\rule{0em}{2.25ex}intuitive picture in the presence of non-zero quark mass.
The 4-flavor scheme component of the general formalism \emph{includes the
full charm quark mass effects, after the infra-red unsafe part has been
resummed}. It represents a substantial improvement over the conventional
4-flavor formalism in the region where $ln(Q^2/m_c^2)$ is not very large,
which includes a substantial fraction of the current experimental range. This
general approach has now been adopted, in different guises, by most recent
papers on heavy quark production in PQCD.
\cite{Kniehl:1995em,BuzaEtal,ThorneRob,CaccGreNas,CaccFrixNas} }
\raisebox{1.5ex}{\parbox{0.4\textwidth}{ \epsfxsize=0.4\textwidth
\epsfbox{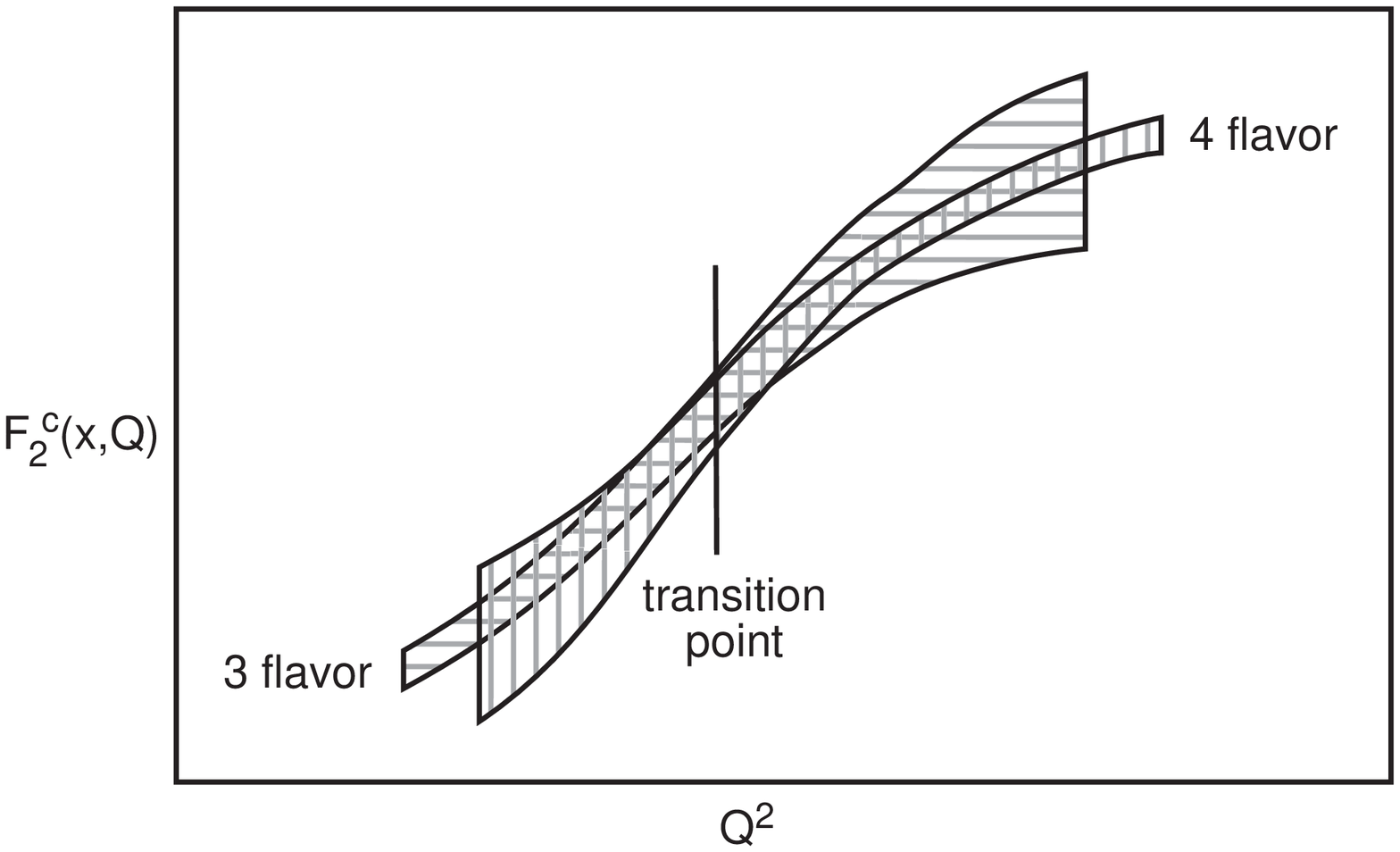} \centering{\footnotesize \textbf{Figure 4.} Intuitive
picture of the \\ general (composite) scheme.} }
}\addtocounter{figure}{1}%

\rule{0em}{2.25ex} The intuitively ``obvious'' general formalism is also
technically precise: the order-by-order rules of calculation can be stated
succinctly \cite{ASTW}; and the validity of the factorization theorem which
underlies it can be established to all orders of perturbation theory
\cite{Collins97}. The essential ingredients of this formalism are:
\\
$\bullet$ \rule{0em}{2.5ex}\underline{\em 3-flavor scheme} at physical scales
$Q \sim m_c$ and extending up;
\\
$\bullet$ \underline{\em 4-flavor $m_c \neq 0$ scheme} at asymptotic $Q \gg
m_c$ and extending down;
\\
$\bullet$ a set of \underline{\em matching conditions} which relate the two
schemes at some
 scale $\mu_m$;
\\
$\bullet$ a suitable \underline{\em transition} scale $\mu_t$ at which one
switches from one scheme to the other.
\\
\rule{0em}{2.5ex} There are considerable inherent flexibility in the choice
of $\mu_m$ and  $\mu_t$, which partially account for the apparent differences
in recent papers on this subject. For detail discussion, see
Refs.\cite{ASTW,Collins97}. We note that this composite scheme is just an
extension of the conventional (zero-mass) variable flavor number scheme
(VFNS), to include heavy quark mass effects in a rigorous way. We will simply
refer to it as the VFNS in the following discussions.
%

%\Input{results1.tex}
%TCIDATA{LaTeXparent=0,0,RingbergProc.tex}

%TCIDATA{ChildDefaults=chapter:3,page:1}

\section{Comparison to Recent HERA Data on Inclusive Charm Production and
Importance of the non-zero-mass 4-flavor Scheme} \label{sec:Result1}

With the use of the general formalism, the theory of inclusive structure
functions, including heavy quark mass effects, is on firm ground. The
comparison of the charm component of this, $F_2^c(x,Q)$, to measurement is,
however, subject to some theoretical and experimental subtleties
\cite{ASTW,ChuvakinEtal}. Previous comparison between the NLO 3-flavor
calculation (of order $\alpha_s^2$ in this scheme) \cite{Harris} with recent
HERA data \cite{ZeusF2c,Redondo:2001iy} showed good agreement. One can also
compare the NLO 4-flavor calculation (of order $\alpha_s$ in this scheme)
with the same data \cite{ASTW}. The results of this calculation are presented
in Fig.~\ref{fig:Five}, along with previous results. We see that the
agreement with data is also excellent. This comparison tells us that, at
least within the current experimental kinematic region, both 3-flavor and
4-flavor schemes are robust, in the sense that both can be applied to the
full range without explicit evidence of the inadequacies expected from the
theoretical considerations described in the previous section. In other words,
both work better than expected. This is not guaranteed to hold indefinitely,
however, for future expanded kinematic ranges. In fact, as we shall see,
calculations show clear discrepancies between the two schemes at moderate $x$
(say between 0.01 and 0.2) and large $Q$. \figFive

It is encouraging that, the NLO 4-flavor (order $\alpha_s$) calculation in
the general formalism appears to maintain good accuracy approaching the
threshold region from above, since it enjoys the advantage of being much
simpler than the 3-flavor NLO (order $\alpha_s^2$) calculation. This is
important for phenomenology.  For instance, in Global QCD Analysis of parton
distributions, the charm contribution to the total inclusive cross-section is
quite significant -- up to 25\% at small x.  Since the total inclusive
structure functions are always treated in the variable flavor number scheme
(VFNS), the charm component is naturally treated the same way.

Furthermore, the 4-flavor scheme is inherently more versatile than the
3-flavor scheme. It can accommodate, in principle, one more non-perturbative
degree of freedom -- a charm component inside the nucleon -- which is
non-existent in the 3-flavor scheme. In view of the dilemma confronting the
phenomenology of bottom production described in the introduction, it is
important to leave open the possibility of unexpected non-perturbative heavy
quark contribution to the structure of the nucleon. Only detailed
phenomenological study done in the VFNS will be able to tell whether such
components actually exist in nature.

These considerations suggest that it is important to examine in more detail
implementation issues of the VFNS schemes (in which heavy quarks participate
as partons) -- issues associated with the choice of prescriptions near the
threshold region, which affect the predictive power of the calculation.

%\Input{implementations.tex}
%TCIDATA{LaTeXparent=0,0,RingbergProc.tex}

%TCIDATA{ChildDefaults=chapter:3,page:1}

\section{Implementation Issues Associated with Variable Flavor Schemes}

\label{sec:ImpleIssues}

In this section, we will quickly summarize the general features of the
various existing calculations of open heavy flavor production in the VFNS,
then focus on the practical issues related to implementation choices. All the
approaches should agree at high energies, within the accuracy of the
perturbative approach. The differences are mainly in the threshold region,
where heavy quark partons are less well-defined (as discussed in
Sec.~\ref{sec:ConvAppr}), hence the perturbation theory contains more
prescription dependence. We will show, however, that other relevant physical
considerations, especially kinematic constraints of the overall heavy flavor
production process, can provide valuable input which considerably improves
the predictive power of the calculation.

The fixed-flavor-number scheme is the scheme of choice in the threshold
region, provided there is no non-perturbative heavy flavor component of the
nucleon. It provides more definitive predictions in this region, and it
respects the heavy quark production kinematics. Hence its results can be used
for comparing and calibrating the different prescriptions for performing
variable flavor calculations. This will guide our analysis which follows.

For simplicity, we continue to restrict ourselves to the case of 4-flavor
scheme calculation of charm production in DIS to order $\alpha_s$. All ideas
apply to higher orders, and to bottom production.

\subsection{General features of the 4-flavor calculation}

\label{sec:GenFeat}

Since the original formulation of (non-zero mass) 4-flavor scheme \cite
{pAcot,ACOT}, a variety of variable flavor number scheme (VFNS) calculations
have been applied to photo- \cite{CaccFrixNas}, lepto- \cite
{Kniehl:1995em,BuzaEtal,ThorneRob,ChuvakinEtal}, and hadro-production
\cite{CaccGreNas} of charm. In spite of apparent differences in formulation
and implementation, the underlying ideas described in Sec.~\ref{sec:ConvAppr}
are adopted by all these calculations. The general formula written down in
\cite{ACOT} which embodies these ideas, has the structure,
\begin{equation}
\hspace{-3.5em}\textrm{[LO\,4-flv scheme\;term]\, -\,
[asymptotic/subtract\;term]\, +\, [LO\,3-flv\;scheme\;term] } \label{eq:a}
\end{equation}
Pictorially, these terms can be represented, for DIS charm production at the
lowest non-trivial order as,\newline \epsfxsize=0.9\textwidth
\centerline{\epsfbox{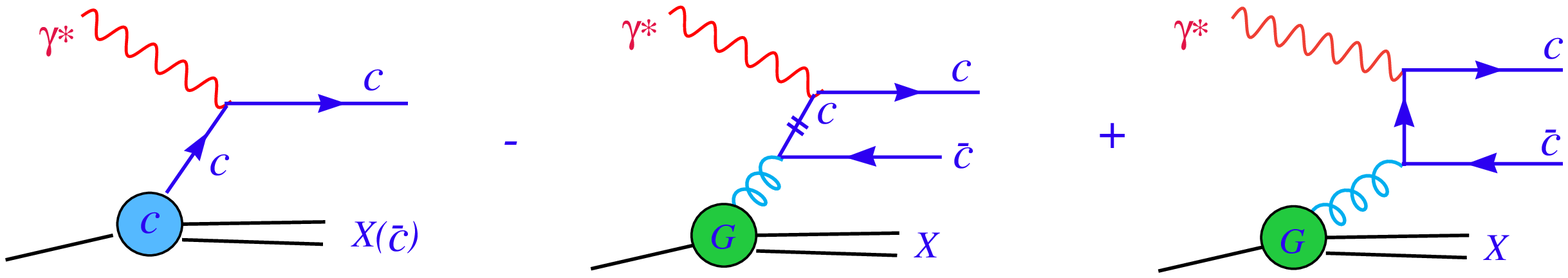}}\newline%
This basic structure appears in all the VFNS calculations in different
guises. The first (LO 4-flv scheme) term is variously called the
flavor-excitation, or quark-scattering, or resummed term -- as the 4-flv
scheme resums the large logarithms associated with the mass of the charm
quark. The third (LO 3-flv scheme) term is variously referred to as the
flavor-creation, or gluon-fusion, or fixed-flavor-number term -- since the
charm quark never becomes an active parton flavor. The middle
(asymptotic/subtract) term represents the overlap between the LO 3-flv scheme
and LO 4-flv scheme terms; hence it needs to be subtracted in order to avoid
double-counting.

At high energies, the asymptotic/subtract term contains the logarithmic mass
singularities of the 3-flv scheme calculation. It constitutes the subtraction
which is needed to make the latter infra-red safe. Together, they form part
of the next-order correction to the dominant resummed LO 4-flv scheme term,
which gives the most accurate physical picture at high energies
(cf.~Sec.~\ref{sec:Intro}). In contrast, near the threshold region, the same
subtraction term overlaps strongly with the first (flavor excitation) term,
since in this region the charm parton arises primarily from a single gluon
splitting, as seen in the pictorial illustration. In this region, the LO
3-flv  scheme (third) term gives the best representation of the correct
physics, as discussed in Sec.~\ref{sec:Intro}. In a consistent application of
the PQCD formalism, the subtraction term is automatically generated by the
renormalization and factorization procedure, as shown in \cite{pAcot,ACOT}.
However, this procedure does not dictate every detail of the implementation.
It allows some degree of prescription-dependence, which is in addition to the
already well-known scheme dependence of massless PQCD.

The prescription-dependence associated with non-zero charm quark mass is most
noticeable at a low energy scale, not far above threshold. In principle, this
dependence can be minimized by choosing the transition scale
$\mu_t$ from the 3-flv scheme to the 4-flv scheme (cf.~Sec.~%
\ref{sec:GenFrame}) to be relatively high. However, in practice, it is
desirable to use the 4-flv scheme even at lower energy scales for reasons
discussed in the introduction and at the end of Sec.\ \ref{sec:Result1}.
Accordingly, we will examine in some detail the implementation issues of the
4-flv scheme, using PQCD as well as other applicable physical constraints,
and see to what extent can its range of predictions be narrowed.

\subsection{Constraints on the Implementation of the 4-flavor scheme}
\label{sec:constraints}

Consider one of the inclusive structure functions in DIS. The factorization
theorem, including quark mass effects \cite{Collins97}, takes the form
\begin{equation}
F(x,Q)=\int \frac{dz }{z }\,f_{a}(z ,\mu )\, \hat{\omega}^a(\frac{x}{%
z},\frac{Q}{\mu},\frac{m_{H}}{\mu},\alpha _{s}(\mu))
\;+\;O(\alpha_{s}^{n+1},\frac{\Lambda^2}{Q^2},\frac{\Lambda^2}{m_{H}^2})
\label{eq:f}
\end{equation}
where $f_{a}$ is the parton distribution function, $\hat{\omega}^{a}$ is the
hard scattering amplitude calculated in PQCD to some power of $\alpha _{s},$
say $n$), and $m_{H}$ is the heavy quark mass. The prescription-dependence
allowed by the PQCD formalism is associated with possible implementations of
the first term of Eq.~\ref{eq:f} within the accuracy specified by the order
of magnitude of the remainder term.

For the case we use as an illustration, the factorization formula consists of
the three terms given in Eq.~\ref{eq:a}. The explicit expressions are:
\begin{equation}
\begin{array}{l}
{\displaystyle c(\zeta ,\mu )\;\omega ^{0}\;-\;\alpha _{s}(\mu )\ln (\frac{%
\mu }{m_{H}})\int_{\zeta }^{1}\frac{dz}{z}g(z,\mu )P_{g\rightarrow c}(\frac{%
\zeta }{z})\;\omega ^{0}} \\
\hspace{4.8em}{\displaystyle +\;\alpha _{s}(\mu )\int_{\chi
}^{1}\frac{dz}{z}g(z,\mu )\omega ^{1}(\frac{\chi
}{z},\frac{m_{H}}{Q})\rule{0em}{3ex}}
\end{array}
\label{eq:b}
\end{equation}
The third, LO 3-flv scheme (gluon fusion), term is well-defined in the
threshold region. The kinematic variable $\chi =x(1+4m_{H}^{2}/Q^{2})$ can be
interpreted as the ``rescaling variable'' for creating a pair of heavy quarks
from massless partons (cf.\ the graph associated with this term in
Eq.~\ref{eq:a}), and the hard cross-section ${\omega ^{1}(\frac{\chi
}{z},\frac{m_{H}}{Q})}$ contains the full heavy quark mass dependence. Note
that, at high $Q$ values ($m_{H}^{2}/Q^{2} \ll 1$), the rescaling variable
reduces to $x$, $\chi \rightarrow x$.

The first two terms of Eq.\ref{eq:b} (the LO resummed and the subtraction
terms) contain the main prescription-dependent mass effects.%
\footnote{These are in addition to the scheme- and scale-dependance known in
the massless
theory.} %
Different versions of the VFNS scheme in the literature mostly amount to
different prescriptions to implement heavy quark mass effects of these terms
or their equivalents (cf.\ Sec.~\ref{sec:compare}). It is therefore important
to examine the origin of the prescription-dependence, and to identify
implementations which are physically reasonable and which lead to numerically
stable results.

Prescription-dependence related to quark mass effect enter Eq.~\ref{eq:b} in
two ways: the so-far unspecified scaling variable $\zeta $, and the
mass-dependence of the LO 4-flv scheme hard cross-section $\omega^0$. From
the PQCD factorization viewpoint, the requirements are that $\zeta
\rightarrow x$ and $\omega^0(m_{H}^{2}/Q^{2})\rightarrow \omega^0(0) $  when
$Q^2 \gg m_H^2$. Thus, at high energies, the charm quark behaves just like a
light quark. At low energies, the exact treatment of $\zeta $ and $\omega^0$
is not prescribed by factorization; however, it makes sense to use the same
$\zeta $ and $\omega^0$ in the first two terms of Eq.~\ref{eq:b}. This
ensures the desired cancellation between these terms (in the absence of
non-perturbative charm), leaving the 3-flavor (gluon fusion) term as the
dominant contribution near the threshold region, as expected from physical
considerations, cf.~\ref{sec:ConvAppr}.

It has been known since the original ACOT paper \cite{ACOT} that the naive
choice of kinematic variable $\zeta$ based on the LO 4-flv scheme $2
\rightarrow 1$ process leads to unnatural results (because kinematical
effects due to the unobserved heavy quark in the target fragment are
ignored). The various proposals to remedy this problem in the literature
either exploit PQCD features (such as ad hoc choices of the scale variable
$\mu$ \cite{ACOT,BuzaEtal,ChuvakinEtal}) or rely on ad hoc threshold
regulating factors \cite{CaccFrixNas,CaccGreNas} or matching conditions
\cite{ThorneRob}.%
\footnote{The sensitivity to the choice of mass-dependence of
$\omega^0(m_{H}^{2}/Q^{2})$ is less significant as that to the choice of
$\zeta$. The easiest choice is to let $\omega^0=\omega^0(m_{H}=0)$
\cite{SACOT}. We shall not discuss this issue in this talk. \label{fn:sacot}}

Since the main arbitrariness of the first two terms in Eq.~\ref{eq:b} is of
kinematic origin, its resolution is most naturally obtained by a kinematic
treatment (so far unexplored). In the next subsection, we show how this can
be easily accomplished. Then, in Sec.~\ref{sec:compare}, we will compare the
results from this treatment with those from the more elaborate ones in the
existing literature.

\subsection{Kinematic solution to the threshold problem}
\label{sec:solution}

\parbox{0.65\textwidth}%
{It is well known that, in a generic process, producing a $H\bar{H}$ pair,
the final-state phase space close to threshold is proportional to $\Delta = 1
- 4m_H^2/W^2$ where $W$ is the CM energy ($W^2=Q^2(x^{-1}-1)$ in DIS). This
factor arises algebraically from the phase space calculation in either of two
ways: (i) in a final transverse momentum integral, it arises from
$\int_0^{p_{CM}^2}\frac{dk_t^2}{k_t^2+m_H^2}|\mathcal{M}|^2$, where $p_{CM}$
is the CM }
\hfill\raisebox{-6.5ex}{\epsfxsize=0.3\textwidth\epsfbox{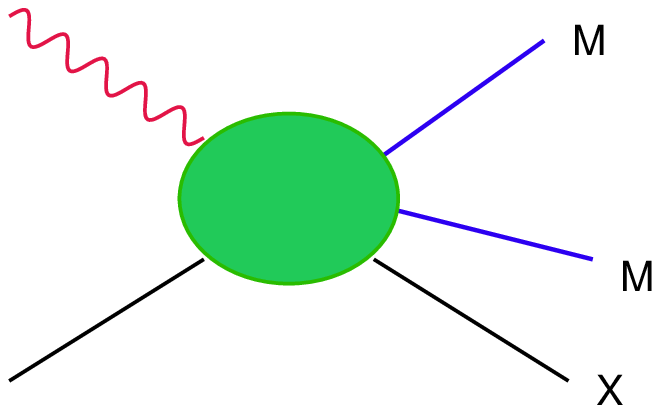}}
3-momentum and $p_{CM}^2 \propto \Delta$; or in a final longitudinal momentum
fraction integral, it arises from $\int_\chi^1
\frac{d\xi}{\xi}|\mathcal{M}|^2$ and the integration range $1-\chi \propto
\Delta$. This generic phase-space factor is a necessary consequence of the
fact that the heavy quark is produced in pairs.

In DIS, $\chi =x(1+4m_{H}^{2}/Q^{2})=1-\Delta(1-x)$; it is the same as the
$\chi$ variable appearing in the gluon fusion term in Eq.~\ref{eq:b}. We see
that $\chi \rightarrow 1$ as $W \rightarrow 2m_H$ (the threshold) from above.
This fact guarantees that the cross-section for the gluon fusion term
vanishes at threshold, as it should. On kinematical grounds, the same
condition should also be satisfied by the first two terms of Eq.~\ref{eq:b}
-- the final state does consist of two heavy quarks, even though only one
explicitly appears in the hard scattering part. We can implement the correct
kinematics for the resummed and the subtraction terms by choosing the
variable $\zeta=\chi$. This choice is also logical from the PQCD point of
view: since the subtraction term represents the part of the gluon fusion term
which is singular at high energies, the integration range is naturally chosen
to be the same to ensure smooth matching.

This simple prescription leads to the following expression for the charm
contribution to the inclusive DIS structure function, from Eq.~\ref{eq:b}:
\begin{equation}
%\begin{array}{l}
\hspace{-5em}
[{\displaystyle c(\chi ,\mu )\;-\;\alpha _{s}(\mu )\ln (\frac{%
\mu }{m_{H}})\int_{\chi }^{1}\frac{dz}{z}g(z,\mu )P_{g\rightarrow c}(\frac{%
\chi }{z})]\;\omega ^{0}}%
{\displaystyle+\alpha _{s}(\mu )\int_{\chi }^{1}\frac{dz}{z}\;g(z,\mu
)\;\omega ^{1}(\frac{\chi }{z},\frac{m_{H}}{Q})\rule{0em}{3ex}}
%\end{array}
\label{eq:c}
\end{equation}
where $\chi =x(1+4m_{H}^{2}/Q^{2})$. We shall call this the ACOT($\chi$)
prescription for implementing the VFNS. In this implementation, the structure
function satisfies the kinematic requirements of charm pair production even
in the presence of non-perturbative charm, provided $ c(\chi ,\mu )
\rightarrow 0$ as $\chi \rightarrow 1$. If there is no non-perturbative
charm, the first two terms will nearly cancel in the threshold region,
leaving the gluon fusion term as the dominant contribution, as expected from
3-flavor calculation.

The use of the rescaling variable $\xi=x(1+m_{H}^{2}/Q^{2})$ for the LO ($s
\rightarrow c$) mechanism in neutrino DIS charm production has been known for
a long time. The use of the $\chi$ rescaling variable for neutral current
$c$-pair-creation discussed here is similar in principle. However, in the
neutrino scattering case, the initial state partons are all light; the
kinematics are obvious. For neutral current scattering, because of the
unobserved heavy quark in the target fragment, the need for a similar
treatment of the kinematics of the ($c \rightarrow c$) term has apparently
been overlooked until now.

Our proposed prescription is in fact applicable beyond the LO flavor
excitation ($c \rightarrow c$) term. We have emphasized the generality of the
kinematic argument, the importance of using the same choice also in the
subtraction term in the NLO implementation, and the relation of this natural
choice to the matching gluon fusion term which contains the correct
kinematics. In fact, although we motivated the ACOT($\chi$) prescription by
the specific example of NLO 4-flv scheme DIS charm production,%
\footnote{The full NLO 4-flv scheme calculation includes, in addition to the
terms shown in Eq.~\ref{eq:a}, an order $\alpha_s$ quark-initiated term with
the associated subtraction term. For simplicity, they have been left out of
the qualitative discussions. They are numerically small in the current
kinematic
region. Nonetheless, they have been included in our calculations shown below.} %
it can be applied to $(N_f+1)\,$-$\,$scheme calculation of heavy quark
production in general. ($N_f$ is the number of light quark flavors.) Roughly
speaking, the rule is: for gluon-initiated subprocesses, use the full $m_H$
kinematics and matrix elements; for heavy-quark initiated subprocesses, use
the rescaling variable appropriate for heavy quark pair production to
restore the correct kinematics.%
\footnote{It is permissible to set $m_H=0$ in the hard matrix element, cf.\
footnote on page \pageref{fn:sacot}.} %

%\Input{compare.tex}
%TCIDATA{LaTeXparent=0,0,RingbergProc.tex}

%TCIDATA{ChildDefaults=chapter:5,page:1}

\section{Results and Comparisons}
\label{sec:compare}

We now show some typical results on the charm contribution to the total
inclusive structure function $F_2(x,Q)$ in the HERA kinematic range.
\figResult Fig.~\ref{fig:Result}a,b compares results obtained using two
different choices of the scale $\mu$. The naive LO 4-flavor result ($\propto
c(x,\mu)$, with Bjorken $x$) is extremely sensitive to the choice of $\mu$.
By itself, it is clearly unphysical in the threshold region. However, once we
adopt the rescaling variable $\chi$ in place of $x$, even this LO 4-flavor
term (labelled LO 4-flv ACOT($\chi$)) behave sensibly compared to the LO
3-flavor and NLO 3-flavor results (which represent the ``right physics'' in
the threshold region, assuming no non-perturbative charm). In addition, the
behavior is much less sensitive to the choice of scale $\mu$, as seen by
comparing the corresponding curves in the two figures.

The original ACOT paper \cite{ACOT} attempted to regulate the threshold
behavior by exploiting the freedom to choose the renormalization and
factorization scale $\mu$. By hindsight, that procedure, as well as others
which rely on ad hoc threshold regularizing schemes
\cite{CaccGreNas,ThorneRob,ChuvakinEtal}, are contrived since they don't
address the underlying kinematic constraint in a direct way. Consequently,
all lead to results which are sensitive to the specifics of the prescription.
As an example, in Fig.~\ref{fig:Result}b, we compare our results with the LO
4-flavor calculation of Thorne-Roberts \cite{ThorneRob} (the curve labelled
$c(\xi)$*TR) which was designed to remedy the threshold behavior of earlier
implementations. This is a rather elaborate scheme, involving lengthy
integral-differential formulas even for the LO term. In the figure, we see
that the threshold behavior of our simple LO term, given by $c(\chi)$, and
that of TR are indistinguishable.  Above threshold, the ACOT($\chi$) curve
naturally interpolates between the correct 3-flavor results at threshold and
the zero-mass 4-flavor ($c(x)$) results at large values of $Q$. The TR curve,
on the other hand, rises steeply above the threshold region, overtaking the
zero-mass 4-flavor result, then approach the same limit from above at high
energies. The sharp rise in the intermediate region appears to be artificial;
and, given the size of the effect, it will have phenomenological
consequences.

Fig.~\ref{fig:ResX} gives an overview of the predictions of the LO and NLO
ACOT($\chi$) calculation compared to the LO and NLO 3-flavor calculation from
$x=10^{-3}$ to $x=0.1$, in the $Q/W$ range of the HERA measurements. As
before, using this prescription, we observe the correct physical threshold
behavior both at LO and at NLO. In addition, the fractional change due to the
NLO correction is relatively small over the full range of $W/Q$; i.e.\ the
perturbative series is ``radiatively stable''. The results are also stable
with respect to the choice of scale $\mu$. More comprehensive results on the
ACOT($\chi$) calculation, and its applications to global analysis and heavy
flavor production will be presented in \cite{KST01}.
\figResX

%\Input{conclude.tex}
%TCIDATA{LaTeXparent=0,0,RingbergProc.tex}

%TCIDATA{ChildDefaults=chapter:6,page:1}

\section{Summary}

We conclude by summarizing the main points of this talk.

%\begin{Simlis}[]{0em}
 The (fixed) 3-flavor scheme calculation for charm production in DIS has
worked well phenomenologically up to now. But, (i) it is inadequate for the
more general purpose of global QCD analysis since other hard processes
require a more general VFNS; (ii) it will become unreliable at some higher
energy scale which cannot be predicted by PQCD; and (iii) it is intrinsically
unequipped to include non-perturbative heavy flavor degrees of freedom should
these occur in nature.

 The (fixed) 4-flavor scheme calculation for bottom production fails
in every experimental comparison performed to date. Among other
possibilities, this could be the smoking gun for non-perturbative effects
beyond fixed order (radiatively-generated) heavy flavor formalisms. This
problem has to be studied in detail, including both charm and bottom.

 In order to study the underlying physics of heavy flavor production in
a systematic way, it is important to have a general PQCD formalism which is
applicable over the full range of energy scales. Such a framework (``VFNS''
with non-zero quark masses) exists \cite{ACOT}, and is on a firm theoretical
basis \cite{Collins97}. It is conceptually simple.  The universal parton
distributions satisfy the usual mass-independent evolution equation. Matching
between different flavor number schemes is known to NLO \cite{ChuvakinEtal}.

 The non-zero mass VFNS provides the possibility of incorporating additional
non-perturbative degrees of freedom associated with heavy quarks. Currently,
there exist several different versions of the VFNS of varying degree of
complexity and naturalness. All implementations approach the same high energy
limit given by the conventional zero-mass PQCD. However, so far, most
prescriptions are sensitive in the threshold region to the ad hoc choices
made in their implementation; some exhibit artificial features in this region
as a result.

 We point out in this talk that the main source of these problems lies in the
lack of proper treatment of kinematics in the resummed term(s) of the VFNS,
due to the neglect of the missing heavy particle in the target fragment. Once
this neglect is remedied (by the use of the natural rescaling variable), a
simple and general method to implement the VFNS emerges. This method yields
amazingly stable and accurate results for charm production already at order
$\alpha_s$ (which is NLO in this scheme) -- remaining uncertainties due to
scale choice and higher-order corrections are shown to be small. We call this
the ACOT($\chi$) scheme. It can provide a practical tool for the systematical
investigation of the non-perturbative heavy flavor degrees of freedom, and
for global QCD analysis with full treatment of heavy-flavor mass effects.

%\end{Simlis}

Details of the material presented here as well as references which cannot be
included in these pages can be found in \cite{ASTW,KST01} and references
cited therein.

\paragraph{Acknowledgement} We would like to thank John Collins and Frederick
Olness for many discussions, and for fruitful collaborations on many aspects
of the subjects discussed in the review part of this talk.

\vspace{3ex}

%\Input{Ringberg.cit}

\end{document}